\documentclass[vecphys]{svmult}
\usepackage{makeidx}
\usepackage{graphicx}
\usepackage{multicol}
\usepackage[bottom]{footmisc}
\makeindex 

\begin{document}

\title*{Thermodynamic curvature and black holes}
\author{George Ruppeiner}
\institute{Division of Natural Sciences, New College of Florida, Sarasota, Florida 34243-2109
\texttt{ruppeiner@ncf.edu}}
\maketitle

In my talk, I will discuss black hole thermodynamics, particularly what happens when you add thermodynamic curvature to the mix. Although black hole thermodynamics is a little off the main theme of this workshop, I hope nevertheless that my message will be of some interest to researchers in supersymmetry and supergravity. Black hole thermodynamics would appear very much in need of some microscopic foundation. We might ask: what are black holes made out of? I will give no answer, but I would like to suggest that what I present here might offer some guidance about the microscopic character of black holes.

\par
Thermodynamic curvature is an element of thermodynamic metric geometry. A pioneering paper on this was by Weinhold \cite{Weinhold1975} who introduced a thermodynamic energy inner product. This led to the work of Ruppeiner \cite{Ruppeiner1979} who wrote a Riemannian thermodynamic entropy metric to represent thermodynamic fluctuation theory, and was the first to systematically calculate the thermodynamic Ricci curvature scalar $R$. A parallel effort was by Andresen, Salamon, and Berry \cite{Andresen1984} who began the systematic application of the thermodynamic entropy metric to characterize finite time thermodynamic processes.

\par
This talk presents a review of thermodynamic curvature $R$ broad in scope, though far from complete in its coverage. I extend the themes discussed in a previous talk \cite{Ruppeiner2013}. My main focus is on achieving some understanding of thermodynamic curvature in the black hole setting. To accomplish this, my working assumption is that for black holes, $R$ follows the same physical interpretation as for ordinary thermodynamic systems, where $R$ gives the size of organized microscopic structures. I present a review of what is known about ordinary thermodynamics, and what this might tell us about black holes.

\section{What is thermodynamic curvature $R$?}
\label{sec:1}

\begin{figure}
\centering
\includegraphics[height=4cm]{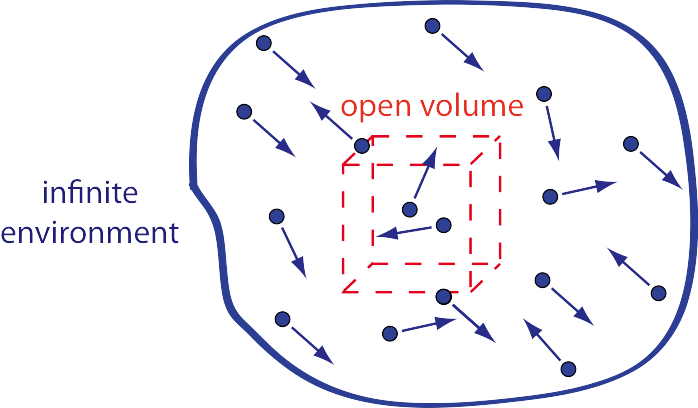}
\caption{An infinite environment of particles and an open volume, with fixed volume $V$, into which particles fluctuate in and out.}
\label{fig:1}
\end{figure}

Thermodynamic curvature comes from thermodynamic fluctuation theory. This classical theory is described in every book on statistical mechanics; it is chapter twelve in Landau and Lifshitz \cite{Landau1977}. For a fluid system, the basic set-up is shown in Figure \ref{fig:1}. There is a infinite universe of particles and some imaginary open volume with fixed volume $V$, into which the particles can travel freely in and out. What is the probability of finding some energy $U$ and some number of particles $N$ in the open volume? Thermodynamic fluctuation theory gives the answer.

\par
Let the particles in the open volume, and the environment consisting of the rest of the particles, be two thermodynamic systems. Denote the fixed thermodynamic state of the environment by ``0''. The thermodynamic state of the open volume fluctuates about an equilibrium characterized by maximum total entropy. The probability of a fluctuation away from equilibrium is given by Einstein's famous Gaussian thermodynamic fluctuation formula \cite{Landau1977, Pathria1996, Einstein1904}:

\begin{equation} \mbox{probability}\propto\mbox{exp}\left[-\frac{V}{2}(\Delta\ell)^2 \right],\label{10} \end{equation}

\noindent where

\begin{equation} (\Delta \ell)^2 = g_{\mu\nu}\Delta x^{\mu}\Delta x^{\nu},\label{20}\end{equation}

\noindent $x^1$ and $x^2$ denote a pair of independent fluctuating thermodynamic variables of the open volume, $\Delta x^{\alpha}= (x^\alpha -x^\alpha_0)$ denotes the difference between $x^{\alpha}$ and its equilibrium value $x^{\alpha}_0$, where the total entropy is maximized, and $g_{\mu\nu}$ denotes the elements of the thermodynamic entropy metric discussed below.

\par Let $S$, $X^1$, and $X^2$ be the entropy, internal energy $U$, and particle number $N$, respectively, of the open volume. Regard $S=S(X^1,X^2,V)$, with $V$ fixed. $X^1$ and $X^2$ correspond to conserved quantities, and $S$ is additive between the open volume and its environment. If $x^{\alpha}=X^{\alpha}$, then

\begin{equation} g_{\alpha\beta} = -\frac{1}{k_B V}\frac{\partial^2 S} {\partial X^{\alpha}\partial X^{\beta}},\label{30}\end{equation}

\noindent where $k_B$ is Boltzmann's constant \cite{Landau1977, Ruppeiner1995}. Since probability depends only on the thermodynamic state, the metric elements $g_{\alpha\beta}$ constitute a second-rank tensor. $g_{\alpha\beta}$ is a positive definite matrix, since the entropy has a maximum value in equilibrium. This is the condition of thermodynamic stability.

\par
This is all found in Landau and Lifshitz \cite{Landau1977}. Let me now get into some things Landau and Lifshitz did not say. The quadratic form $(\Delta \ell)^2$ in Eq. (\ref{20}) has the look of a distance between thermodynamic states, a distance in the form of a Riemannian metric. The physical interpretation is that: the less the probability of a fluctuation between two states, the further apart they are.

\par
A Riemannian metric in any context leads directly to a Ricci curvature scalar $R$ \cite{Laugwitz1965}, and this is certainly the case here. $R$ is the only geometric scalar invariant function in thermodynamics, and so it must be very fundamental. The units of the thermodynamic curvature are those of volume per particle, and this limits its possible physical interpretation greatly. Units alone suggest that $R$ is a measure of the characteristic size of some sort of organized fluctuating structures within the system.

\par
$R$ is readily calculable from the thermodynamic metric elements $g_{\alpha\beta}$. For example, in $(x^1,x^2)=(T,\rho)$ coordinates, where $T$ is the temperature and $\rho$ is the particle number density, we have the Helmholtz free energy per volume $f=f(T,\rho)$, the entropy per volume $s=-f_{,T}$ (where the comma notation indicates partial differentiation), and the chemical potential $\mu=f_{,\rho}$. The diagonal metric elements ($g_{12}=0$) are \cite{Ruppeiner2010}

\begin{equation} g_{11} =\frac{1}{k_B T}\left(\frac{\partial s} {\partial T}\right)_{\rho}, \label{40}\end{equation}

\noindent and

\begin{equation} g_{22} =\frac{1}{k_B T}\left(\frac{\partial \mu} {\partial \rho}\right)_{T}. \label{50}\end{equation}

\noindent For a diagonal metric \cite{Laugwitz1965}

\begin{equation}
R=\frac{1}{\sqrt{g}}\left[\frac{\partial}{\partial x^1} \left(\frac{1}{\sqrt{g}}\frac{\partial g_{22}}{\partial x^1}\right)+\frac{\partial}{\partial x^2}\left(\frac{1}{\sqrt{g}}\frac{\partial g_{11}}{\partial x^2}\right)\right],\label{60}
\end{equation}

\noindent where

\begin{equation} g=g_{11}\,g_{22}.\label{70} \end{equation}

\par
A simple example is the ideal gas, in which there is no interaction between the particles. Here

\begin{equation} f(T,\rho)=\rho k_B T\, \mbox{ln}\rho + \rho k_B h(T), \label{80}\end{equation}

\noindent where $h(T)$ is some function of the temperature with negative second derivative. Eq. (\ref{60}) now yields $R=0$ \cite{Ruppeiner1979}. This suggests that $R$ is some type of measure of interactions between particles.

\par
Calculations in critical point models show that $|R|$ diverges as the correlation volume $\xi^d$, where $d$ is the spatial dimension of the system \cite{Ruppeiner1979, Ruppeiner1995, Johnston2003}. The connection of $|R|$ to fluctuating structure size has also been established directly by means of a covariant thermodynamic fluctuation theory \cite{Ruppeiner1995, Ruppeiner2010, Ruppeiner1983a, Ruppeiner1983b, Diosi1985}.

\par
\begin{figure}
\centering
\includegraphics[height=4cm]{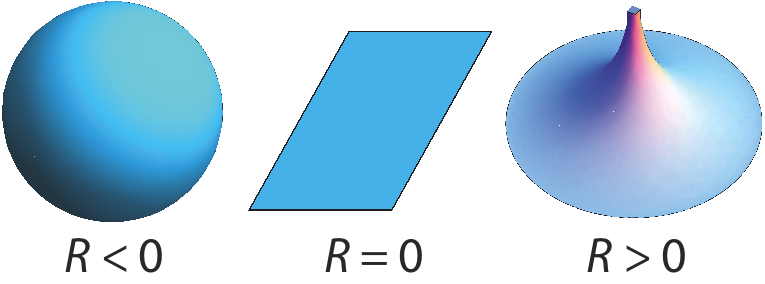}
\caption{Three surfaces with constant Ricci curvature scalar $R$: the sphere, the plane, and the pseudosphere. For pure fluids, in Weinberg's sign convention, $R<0$ if attractive interparticle interactions dominate, and $R>0$ if repulsive interactions dominate. Regardless the sign convention for $R$, attractive interactions correspond to the geometry of the sphere, and repulsive interactions to the geometry of the pseudosphere.}
\label{fig:2}
\end{figure}

$R$ is a signed quantity, as shown in Figure \ref{fig:2}. I use the sign convention of Weinberg \cite{Weinberg1972}. (Sign conventions differ among authors. I express all results reported here in Weinberg's sign convention.) For fluid and solid systems, an overall pattern is that $R$ is negative for systems where attractive interparticle interactions dominate, and positive where repulsive interactions dominate. The sign of $R$ alone thus offers direct information about the character of the interactions among the particles.

\section{$R$ for ordinary thermodynamics}
\label{sec:2}

$R$ has been worked out in a number of cases in ordinary thermodynamics. On systematic tabulation, patterns readily become evident. Such patterns might lend insight into the nature of black hole microscopic properties.

\par
In this section I attempt a classification of the ``basic food groups'' of $R$ for ordinary thermodynamics. Thermodynamics divides neatly into atomic and molecular systems, like fluids and solids, and discrete lattice systems, like magnetic spin systems. I will treat them separately.

\subsection{$R$ for fluid and solid systems, basic models}
\label{sec:21}

In this subsection, I tabulate results for fluid and solid systems, including the quantum gasses. I pay special attention to Lennard-Jones type interacting systems, for which there are a number of interesting recent results.

\par
Table \ref{tab:1} shows $R$ for a number of simple models for which $R$ has only one sign. These models were worked out by a number of authors over a period of years. In these models interactions between particles may take place by virtue of a potential between the particles, or through quantum statistics. In either case, particles tend to either bunch together (attract) or to push apart (repel) compared with the ideal gas. The results in Table \ref{tab:1} clearly show the relation between the character of the interparticle interactions and the sign of $R$. If interactions between particles are attractive, $R$ is negative. Prominent here is the ideal Bose gas,\footnote{The calculation of $R$ for the ideal Bose gas was done with a continuous density of states, and so a possible divergence of $R$ at a Bose-Einstein phase transition with $T>0$ would not have been revealed.} and the typical critical point models. If interactions are repulsive, $R$ is positive. Prominent examples are ideal Fermi gasses. In systems with weak interactions, $|R|$ is zero or ``small,'' where ``small'' means on the order of the molecular volume $v$, $|R|\sim v$ or smaller. Cases with $|R|\sim v$ are typical also of systems dominated by strong short-range repulsive interactions, such as dense liquids and solids. Table \ref{tab:1} also shows where $|R|$ diverges, typically either at absolute zero or at critical points.

\begin{table}
\begin{center}
\begin{tabular}{l|c|c|c|c}
\hline
System															&$n$	& $d$		& \,\,$R$ sign\,\,	& $ \,\, |R|$ divergence\,\,	\\
\hline
Ideal Bose gas				\cite{Mrugala1990,Oshima1999}				& $2$	& $3$		& $-$		& $T\rightarrow 0$		\\
q-deformed bosons			\cite{Mirza2011}							& $2$	& $3$		& $-$		& critical line			\\
Critical regime				\cite{Ruppeiner1979,Ruppeiner1995,Brody1995}	& $2$	& $\cdots$	& $-$		& critical point			\\
Mean-field theory			\cite{Janyszek1989} 							& $2$	& $\cdots$	& $-$		& critical point			\\
van der Waals (critical regime)	\cite{Ruppeiner1995,Brody1995,Brody2009}		& $2$	& $3$ 		& $-$		& critical point			\\
Spherical model			\cite{Johnston2003,Janke2003}				& $2$	& $3$		& $-$		& critical point			\\
Tonks gas					\cite{Ruppeiner1990a}						& $2$	& $1$		& $-$		& $|R|$ small			\\
Ideal gas					\cite{Ruppeiner1979, Nulton1985}				& $2$	& $3$		& $0$		& $|R|$ small			\\
Multicomponent ideal gas		\cite{Ruppeiner1990b}						& $>2$	& $3$		& $+$		& $|R|$ small			\\
Ideal gas paramagnet		\cite{Kaviani1999}							& $3$	& $3$		& $+$		& $|R|$ small			\\
q-deformed fermions		\cite{Mirza2011}							& $2$	& $3$		& $+$		& $T\rightarrow 0$		\\
Ideal Fermi gas				\cite{Mrugala1990,Oshima1999,Ruppeiner2008}	& $2$	& $2,3$		& $+$		& $T\rightarrow 0$		\\
Ideal gas Fermi paramagnet	\cite{Kaviani1999}							& $3$	& $3$		& $+$		& $T\rightarrow 0$		\\
\hline
\end{tabular}
\end{center}
\caption {The thermodynamic curvature $R$ for a number of simple models for which $R$ has only one sign. Tabulated are the number of independent thermodynamic parameters $n$, the spatial dimension $d$, the sign of $R$, and the possible divergences of $R$. For some models, there is no particular spatial dimension $d$, and this is denoted by $\cdots$. ``$|R|$ small'' means that the value of $|R|$ is on the order of the volume of an interparticle spacing or less.}
\label{tab:1}
\end{table}

\par
Table \ref{tab:2} shows four additional models, each having $R$ with both signs. The Takahashi gas has negative $R$ for the gas-like phase, where attractive interactions dominate, and small $|R|$ in the liquid-like phase. Increasing the density at constant low temperature yields a pseudophase transition from a gas-like phase to a liquid-like phase. This pseudophase transition is accompanied by a sharp {\it positive} spike in $R$. Conceptually simpler than the Takahashi gas are the remaining three models in Table \ref{tab:2}, which are all quantum gasses intermediate between Fermions and Bosons, and with sign of $R$ switching from positive to negative through $R=0$ on transitioning from Fermionic to Bosonic behavior. Gentile statistics have an integer parameter $p$ giving the maximum occupation number of a state, with $p=1$ corresponding to a pure Fermi gas, and $p\rightarrow\infty$ to a pure Bose gas. In the same spirit is the $M$-statistics model, with state occupation number $M$. For any temperature and chemical potential, $R$ eventually transitions in sign from positive to negative as $M$ increases from $1$. $R$ thus offers a convincing method of determining when the $M$-statistics model transitions from Fermionic to Bosonic. The quantum gas of anyons is intrinsically two-dimensional, and has particles with fractional spin $\alpha$ whose variation allows us to change it continuously from a Bose gas to a Fermi gas ($\alpha: 0 \rightarrow 1$); the sign of $R$ changes correspondingly from negative to positive.

\begin{table}
\begin{center}
\begin{tabular}{l|c|c|c|c|c}
\hline
System															& \,\,$n\,\,$	&\,\,$d$\,\,		& \,\, $R$ sign	\,\,	&\,\, $R=0 \,\,$	& \,\,$|R|$ divergence\,\,	\\
\hline
Takahashi gas				\cite{Ruppeiner1990a}						& $2$	& $1$		& $\pm$			& Yes		& $T\rightarrow 0$	\\
Gentile's statistics			\cite{Oshima1999}							&$2$	&$3$		&$\pm$			& Yes		& $T\rightarrow 0$	\\
$M$-statistics				\cite{Ubriaco2013}							& $2$	& $ 2, 3 $		& $\pm$			& Yes		& $T\rightarrow 0$	\\
Anyons					\cite{Mirza2009}							& $2$	& $2$		& $\pm$			& Yes		& $T\rightarrow 0$	\\
\hline
\end{tabular}
\end{center}
\caption {The thermodynamic curvature $R$ for models where $R$ has both signs. In each case, $R$ changes sign through $R=0$.}
\label{tab:2}
\end{table}

\subsection{$R$ for fluid and solid systems, Lennard-Jones potential}
\label{sec:22}

\begin{figure}
\centering
\includegraphics[height=4cm]{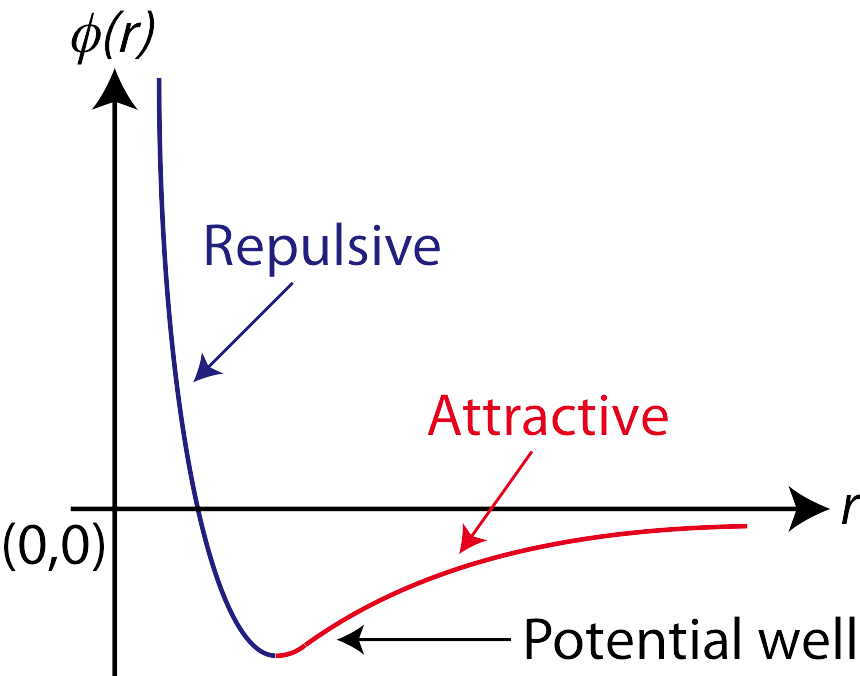}
\caption{The Lennard-Jones type potential, in which two particles separated by a distance $r$ experience a potential $\phi (r)$, repulsive at short range and attractive at long range.}
\label{fig:3}
\end{figure}

A major element in the study of fluid and solid systems is the Lennard-Jones type potential between particles, shown schematically in Figure \ref{fig:3}. This potential approximates the interaction between particles in real fluids and solids. The Lennard-Jones type potential is strongly repulsive at short range and weakly attractive at long range. There is a minimum in the potential where repulsion and attraction balance, and where particles in a condensed liquid or solid phase like to reside. Fluid phases typically posses average separation distances between particles greater than that corresponding to the bottom of the potential well, and so the attractive part of the potential usually dominates. Hence, $R$ should be mostly negative for real pure fluids, which is indeed the case. The study of the Lennard-Jones type interaction supplements that for the simple models above, and takes us a long way towards completing the picture for $R$ for fluid and solid systems.

\par
\begin{figure}
\begin{minipage}[b]{0.5\linewidth}
\includegraphics[width=2.2in]{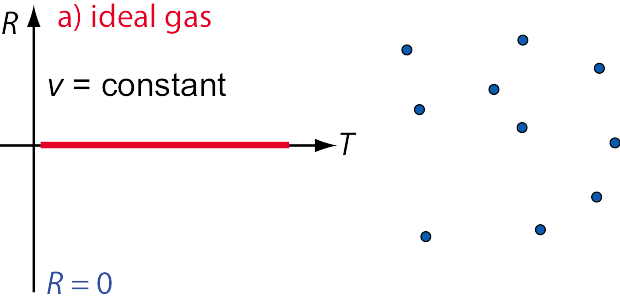}
\end{minipage}
\hspace{0.0cm}
\begin{minipage}[b]{0.5\linewidth}
\includegraphics[width=2.2in]{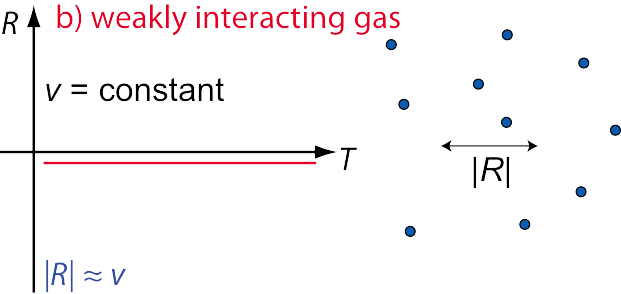}
\end{minipage}
\vspace{0.1cm}
\begin{minipage}[b]{0.5\linewidth}
\includegraphics[width=2.2in]{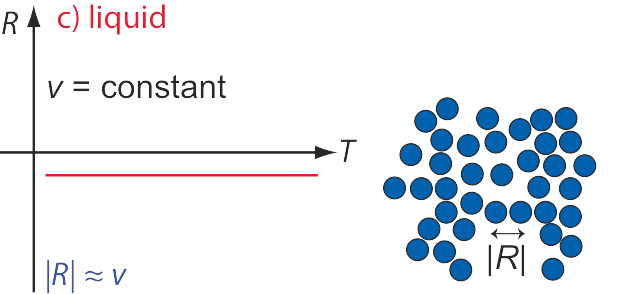}
\end{minipage}
\hspace{0.0cm}
\begin{minipage}[b]{0.5\linewidth}
\includegraphics[width=2.2in]{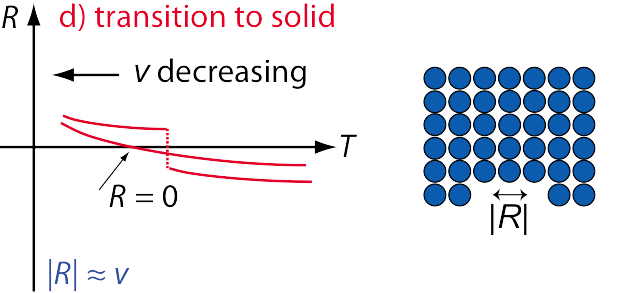}
\end{minipage}
\vspace{0.1cm}
\begin{minipage}[b]{0.5\linewidth}
\includegraphics[width=2.2in]{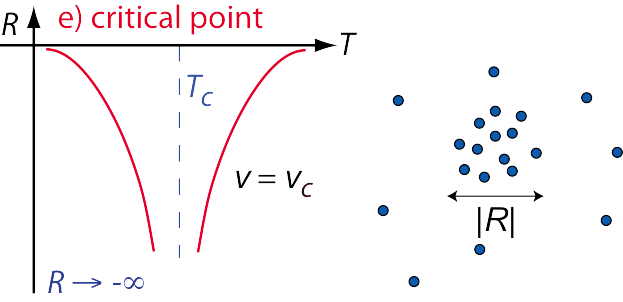}
\end{minipage}
\hspace{0.0cm}
\begin{minipage}[b]{0.5\linewidth}
\includegraphics[width=2.2in]{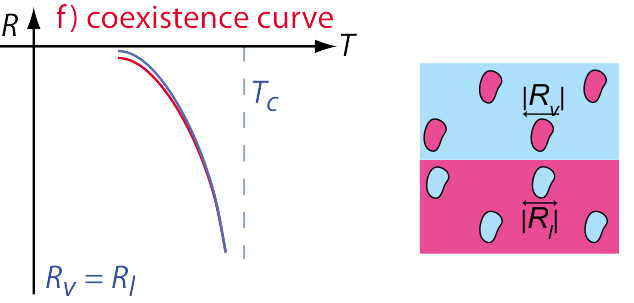}
\end{minipage}
\vspace{0.1cm}
\begin{minipage}[b]{0.5\linewidth}
\includegraphics[width=2.2in]{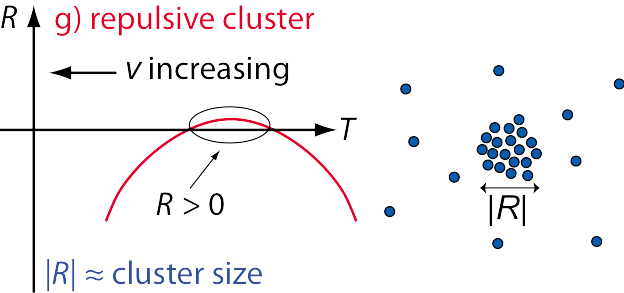}
\end{minipage}
\hspace{0.0cm}
\begin{minipage}[b]{0.5\linewidth}
\includegraphics[width=2.2in]{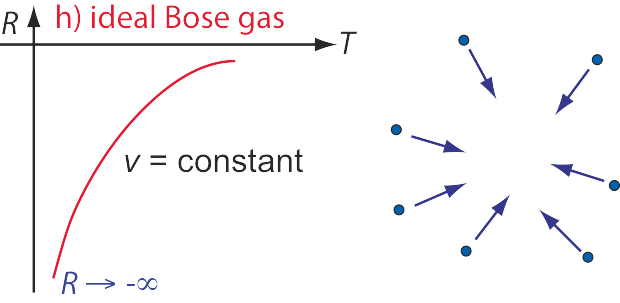}
\end{minipage}
\vspace{0.1cm}
\begin{minipage}[b]{0.5\linewidth}
\includegraphics[width=2.2in]{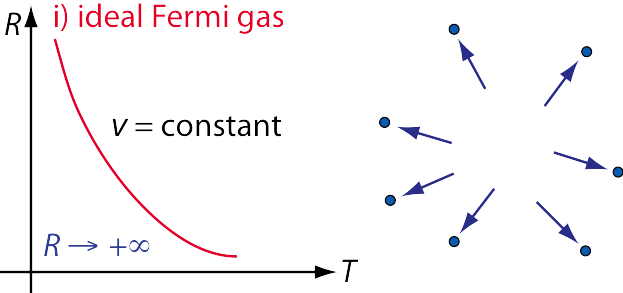}
\end{minipage}
\hspace{0.0cm}
\begin{minipage}[b]{0.5\linewidth}
\includegraphics[width=2.2in]{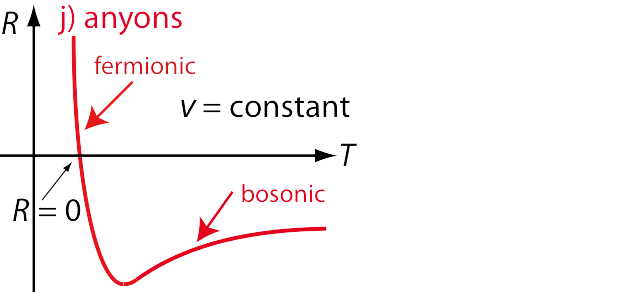}
\end{minipage}
\caption{Schematic graphs for $R$ and corresponding particle configurations: (a) the ideal gas, with $R=0$; (b) the weakly interacting gas, with negative $R$ and $0<|R|\ll v$, where $v$ is the molecular volume; (c) the liquid, with negative $R$ and $|R|\sim v$; (d) the transition from liquid to solid, with $R$ changing sign to positive in the solid, typically discontinuously; (e) the critical point, with $R\rightarrow -\infty$ and $|R|\sim\xi^3$; (f) the coexisting gas and liquid phases, with $R$ equal in the vapor and the liquid phases very near the critical point, $R_l=R_v$; (g) an organized compact repulsive cluster held up by the repulsive part of the interparticle interactions, with positive $R$ and $|R|\sim$ cluster size; (h) the ideal Bose gas, with $R\rightarrow -\infty$ as $T\rightarrow 0$; (i) the ideal 2D or 3D Fermi gas, with $R\rightarrow +\infty$ as $T\rightarrow 0$; and (j) the anyon gas, with a transition from Bose to Fermi behavior as $T$ decreases at fixed $v$.}
\label{fig:4}
\end{figure}

Let me present results from fluid studies based on experimental fluid data \cite{Ruppeiner2012a, Ruppeiner2012b}, and on computer simulations in fluids and solids on particles interacting via an actual Lennard-Jones potential \cite{May2012, May2013}. In each case $R$ was determined by Eq. (\ref{60}), differentiating $f(T,\rho)$ obtained from fits to numerical experimental or computer data. Results of this effort, and the results for the simple models shown above, are summarized in Figure \ref{fig:4}. Figure \ref{fig:4} shows schematic graphs of $R$ as a function of $T$ along curves with the specified $v$. Particle configurations corresponding to each situation are also shown alongside the schematic graphs.

\par
Fig. \ref{fig:4}a shows the ideal gas, which has $R=0$. Fig. \ref{fig:4}b shows the behavior of $R$ perhaps more typical of weakly interacting systems. Here, widely spaced particles interact via the attractive tail of the Lennard-Jones potential. Typically $R$ is negative, and $0<|R|\ll v$. I characterize such situations as having ``small'' $|R|$, even in cases such as near ideal gases where $v$ might get very large. The idea is that at size scales of one molecular volume, the system gets ``grainy,'' and thermodynamic properties such as $R$ based on averages have increasing difficulty being accurate.

\par
The liquid state is shown in Fig. \ref{fig:4}c. We have a compactly arranged, disorganized system of particles held together by attractive interactions, and with negative $R$, and $|R|\sim v$. On compressing the liquid state, there is the possibility of the system organizing into a crystalline solid state, where the predominant interaction is repulsive in character, with $R$ changing sign to positive, and $|R|\sim v$, as shown in Fig. \ref{fig:4}d. Typical is a discontinuous jump from the liquid into the solid state \cite{May2013}.

\par
An essential case is the critical point regime, with $R$ shown in Fig. \ref{fig:4}e. There are two curves for $R$, separated by the critical temperature $T_c$. The curve at lower temperature represents $R$ along the coexistence curve for both liquid and vapor phases, and the curve at higher temperature represents $R$ along the critical isochore $v=v_c$, where $v_c$ is the critical molar volume. $R$ diverges to negative infinity at the critical point along both curves. On the right side of Fig. \ref{fig:4}e, I sketch a near critical point particle configuration where a loose cluster has been formed by the attractive long-range tail of the Lennard-Jones type potential. The size of this cluster is given by the correlation length $\xi$, with $|R|\sim\xi^3$. Another critical point theme is shown in Fig. \ref{fig:4}f, where we have equal values of $R$ for the coexisting liquid and vapor phases, $R_l=R_v$, as the two phases have identical organized droplet sizes \cite{Ruppeiner2012a,Ruppeiner2012b,May2012}.

\par
Fig. \ref{fig:4}g shows a somewhat subtle vapor phase theme \cite{Ruppeiner2012a}. Attractive interactions have formed a tight cluster of solid, which is then pressed together by impacts from surrounding particles. Repulsive interactions hold the structure up and $R$ is positive, with $|R|\sim$ cluster size. Such clusters have been reported in computer simulations in the vapor phase of Water near the critical point \cite{Johansson2005}.

\par
Fig. \ref{fig:4}h shows the ideal Bose gas, with $R$ always negative, and with $R$ diverging to negative infinity as $T\rightarrow 0$. Fig. \ref{fig:4}i shows the ideal Fermi gas, with $R$ always positive, and with $R$ diverging to positive infinity as $T\rightarrow 0$. The ideal Fermi gas shows the same qualitative behavior in 3D \cite{Mrugala1990,Oshima1999} or 2D \cite{Ruppeiner2008}. Fig. \ref{fig:4}j shows the gas of anyons with $0<\alpha<1$. As we cool at constant $v$, starting from a high $T$, $R$ starts with the Bosonic negative sign, but eventually there is a transition to the Fermionic positive sign. Aside from its intrinsic interest, the natural spatial dimension, two, of the anyon gas matches the dimension of black hole event horizons.

\par
\begin{figure}
\centering
\includegraphics[height=5cm]{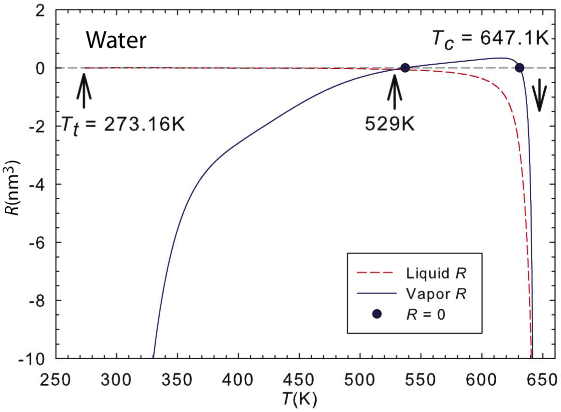}
\caption{$R$ for Water in the coexisting liquid and vapor phases from the triple point to the critical point. Demonstrated are the points made in Figs. \ref{fig:4}b-g.}
\label{fig:5}
\end{figure}

Lest the reader think that this is all theoretical, I show $R$ for Water in Figure \ref{fig:5}, along the coexistence curve in both the liquid and vapor phases. Fig. \ref{fig:5} was worked out with data from the NIST Chemistry WebBook \cite{NIST,Wagner2002}. $R$ is in units of cubic nanometers, and is shown from the critical point $T=T_c$ to the triple point $T=T_t$, where $T_t$ is the triple point temperature. Demonstrated are a number of the principles sketched in Fig. \ref{fig:4}. The predominant sign of $R$ is negative, as the attractive tail of the Lennard-Jones type potential dominates in the fluid.

\par
In conclusion, for fluid and solid systems major elements of the thermodynamic curvature seem to be understood in principle, at least for cases with $n=2$ independent thermodynamic parameters. Cases with $n>2$, such as fluid mixtures, are largely unexplored.

\subsection{$R$ for discrete systems}
\label{sec:23}

The thermodynamic curvature for discrete systems has been less investigated. Spin systems with ferromagnetic interactions tend to have {\it aligned} adjacent spins, and to have critical point properties analogous to those for fluid systems. Indeed, $R$ tends to be nicely negative for ferromagnetic spin systems, with $|R|\sim\xi^d$. By analogy with the fluid systems, then, we might think of ferromagnetic interactions as somehow ``attractive.'' We might also logically think of antiferromagnetic interactions, which tend to {\it disalign} adjacent spins, as ``repulsive,'' and with positive $R$. But there is little evidence that it works out like this. Mirza and Talaei \cite{Mirza2013} worked out $R$ for a model with {\it frustrated} spins, and found a regime with large positive $R$. Perhaps the presence or absence of frustration is the key to interpreting the sign of $R$ for spin systems. More calculations in spin systems would appear indicated before any definitive judgement could be made.

\par
For spin systems, we commonly have a temperature $T$ and a magnetic field $H$ (more than one magnetic field may be present, but this possibility is not explored here). For such models, the partition function gets worked out in terms of $\beta=1/T$ and $h=-H/T$; namely, $Z=Z(\beta,h)$. The partition function leads to the thermodynamic potential per spin $\phi(\beta,h)=\mbox{ln}\,Z$, and the metric elements $g_{\alpha\beta}=\phi_{,\alpha\beta}$, in coordinates $(x^1,x^2)=(\beta,h)$ \cite{Ruppeiner1995}. Here, we set $k_B=1$. It is fashionable in magnetic models to write $R$ as

\begin{equation} R=\frac{\left| \begin{array}{ccc} \phi_{,11}& \phi_{,12}& \phi_{,22}\\ \phi_{,111}&\phi_{,112}&\phi_{,122}\\ \phi_{,112}&\phi_{,122}&\phi_{,222} \end{array}\right|}{2 \left| \begin{array}{cc} \phi_{,11} & \phi_{,12}\\\phi_{,12}&\phi_{,22}\end{array}\right|^2}. \label{90}\end{equation}

\noindent Many of the results for discrete models were worked out with this formula.

\par
Table \ref{tab:3} lists $R$ for some spin models simple enough that $R$ has only one sign (or $R=0$). The first three models in Table \ref{tab:3} have ferromagnetic nearest neighbor interactions. For these, $R$ is negative, with a divergence $R\rightarrow -\infty$ either as $T\rightarrow 0$ (for $d=1$), or at a critical point with $T >0$ (for $d\ne 1$), as interspin coupling brings about a long-range ordering of aligned spins. This situation would appear analogous to the fluid critical point regime. The Ising antiferromagnet has a negative $R$ with magnitude of the order of a lattice spacing, and is similar in this sense to the liquid state of the previous section.

\begin{table}
\begin{center}
\begin{tabular}{l|c|c|c|c}
\hline
System												& \,\,$n\,\,$		& \,\,$d$\,\,		& \,\, $R$ sign\,\,	& \,\,$|R|$ divergence\,\,	\\
\hline
Ising ferromagnet		\cite{Janyszek1989,Ruppeiner1981}	& $2$		& $1$		& $-$			& $T\rightarrow 0$		\\
Ising on Bethe lattice	\cite{Dolan1998}					& $2$		& $\cdots$	& $-$			& critical point			\\
Ising on random graph 	\cite{Johnston2003,Janke2002}		& $2$		& $2$		& $-$			& critical point			\\
Ising antiferromagnet	\cite{Janyszek1989,Ruppeiner1981}	& $2$		& $1$		& $-$			& $|R|$ small			\\
Ideal paramagnet		\cite{Janyszek1989,Ruppeiner1981}	& $2$		& $\cdots$	& $0$			& $|R|$ small			\\
\hline
\end{tabular}
\end{center}
\caption {The thermodynamic curvature $R$ for several simple spin models for which $R$ has only one sign.}
\label{tab:3}
\end{table}

\par
Table \ref{tab:4} shows four discrete systems for which $R$ has both signs. The sign of $R$ for the one-dimensional $q$-state Potts model is related to the number of states per spin $q$. For $q>2$, and nonzero magnetic field, there are significant regimes of positive $R$ at low temperature. An abrupt change in the sign of $R$ is present in the one-dimensional Ising ferromagnet of {\it finite} $N$ spins. $R$ is appropriately negative for large $N$, but sharply increases to large positive values as $N$ is decreased through a volume $N^*\sim |R(N\rightarrow\infty)|$. Work calculating $R$ is in progress for the one-dimensional Ising-Heisenberg model, which shows ferromagnetism, antiferromagnetism, ferrimagnetism, and frustration. The ferrimagnetic phases show substantial regimes of positive $R$. $R$ for the kagome Ising lattice has recently been worked out, mostly in zero magnetic field. This model has a critical line $T=T_c(H)$ in $(T,H)$ space along which $R$ diverges on both sides, negative on the high $T$ side with dominant ferromagnetic interactions, and positive on the low $T$ side with dominant ferrimagnetic interactions.

\begin{table}
\begin{center}
\begin{tabular}{l|c|c|c|c|c}
\hline
System												&\,\,$n$\,\,		&\,\,$d$\,\,		& \,\, $R$ sign\,\,	& \,\,$R=0$\,\,	& $|R|$ divergence	\\
\hline
Potts model $(q>2)$		\cite{Johnston2003,Dolan2002}		& $2$		& $1$		& $\pm$			& Yes		& $T\rightarrow 0$	\\
Finite Ising ferromagnet	\cite{Brody2003}					& $2$		& $1$		& $\pm$			& Yes		& $T\rightarrow 0$	\\
Ising-Heisenberg		\cite{Bellucci2013}					& $2$		& $1$		& $\pm$			& Yes		& $T\rightarrow 0$	\\
Kagome Ising lattice		\cite{Mirza2013}					& $2$		& $2$		& $\pm$			& No			& critical line		\\
\hline
\end{tabular}
\end{center}
\caption {The thermodynamic curvature $R$ for discrete models for which $R$ has both signs.}
\label{tab:4}
\end{table}

\par
The physical interpretation of $R$ for discrete systems is less conclusive than that for the fluid and solid systems. More worked examples are clearly necessary.

\section{$R$ for black hole thermodynamics}
\label{sec:4}

This section discusses black hole thermodynamics, mostly in the context of general relativity \cite{Bekenstein1980}. String theory and other quantum black holes are beyond the scope of this talk.

\subsection{Introduction}
\label{sec:41}

The classical (nonquantum) properties of black holes date to Schwarzchild's solution of Einstein's field equations \cite{Misner1973}. This solution obtains on assuming a static, charge free, spherically symmetric point mass $M$, located at a central singularity. The solution yields a spherical event horizon, centered on the mass, and with radius $r=2M$ (in geometrized units). This event horizon bounds an interior from which there may be no escape, even by light. Einstein's field equations may also be solved if we add charge $Q$ (the Reissner-Nordstr$\ddot{\mbox{o}}$m solution), angular momentum $J$ (the Kerr solution), or if we have all three quantities $(M,J,Q)$ (the Kerr-Newman solution). Hawking, Penrose, and others proved the celebrated uniqueness theorems, that if the collapsing matter is dense enough, then we inevitably approach one of these solutions.

\par
\begin{figure}
\centering
\includegraphics[height=4cm]{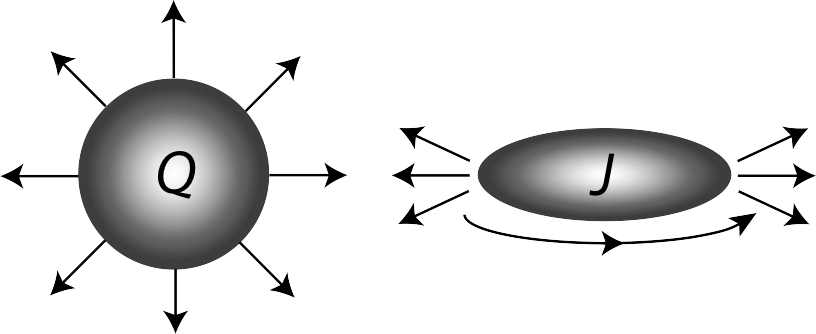}
\caption{Extremal black holes have so much charge $Q$ that they are on the verge of exploding out under their electrostatic repulsion, or so much angular momentum $J$ that they are on the verge of spinning apart. Both of these scenarios, or any combination of them, are forbidden by cosmic censorship.}
\label{fig:6}
\end{figure}

Frequently discussed is the idea of extremal black holes. Could we add enough charge to a black hole so that it explodes outward under its electrostatic repulsion, as in Figure \ref{fig:6}? Or could we add enough angular momentum so that it tears apart under its spin? Cosmic censorship forbids both these scenarios, or any combination of them. We refer to black holes as extremal if they are as close as possible to these mechanical limits. For the Kerr-Newman black hole, the condition of mechanical stability is

\begin{equation} M^4-J^2-M^2 Q^2>0. \label{100}\end{equation}

\noindent At the extremal limit, the black hole temperature $T=0$, and cosmic censorship is a way of expressing the unattainability of absolute zero temperature. It is important to appreciate, however, that the third law of thermodynamics will not always hold for black holes, as extremal black holes do not always have zero area. Hence, the black hole entropy does not always go to zero at zero temperature. This marks an important difference between black hole thermodynamics and ordinary thermodynamics.

\par
\begin{figure}
\centering
\includegraphics[height=5cm]{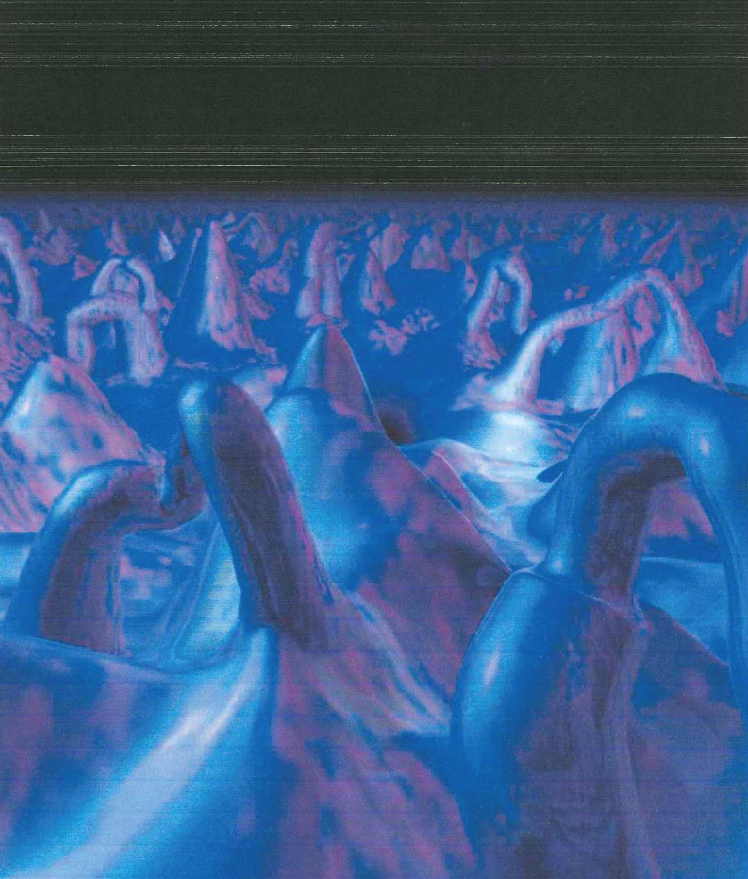}
\caption{Fluctuating event horizon. Are particles associated with these fluctuations? }
\label{fig:7}
\end{figure}

An oft quoted principle of black holes is the ``no-hair theorem'' \cite{Misner1973}. After matter collapses to form a black hole, there is a brief period of settling down during which the history of the black hole's creation is forgotten. The final equilibrium state depends only on $(M,J,Q)$. Such a reduction of complexity is essential for black hole thermodynamics. Taken to its logical extreme, however, and the no-hair conjecture denies the possibility of any form of a distribution of equilibrium black hole microstates. A distribution of microstates is central to statistical mechanics, as well as for thermodynamic fluctuations, and thermodynamic fluctuations are arguably logically necessary to any thermodynamic formalism \cite{Lewis1931}. My working assumption is then certainly to consider fluctuations about the black hole equilibrium thermodynamic state. If we magnify the regime around the black hole event horizon we might expect to see a fluctuating structure perhaps like in Figure \ref{fig:7}. And if we have fluctuations in some quantum structure, would there not be associated fundamental particles?

\subsection{Kerr-Newman black hole thermodynamics}
\label{sec:42}

\par
Consider now the Kerr-Newman black hole thermodynamics, beginning with a comparison to pure fluids; see Table \ref{tab:5}. First, I identify the conserved variables; these play a special role in thermodynamic fluctuation theory \cite{Ruppeiner2007}. For Kerr-Newman black hole thermodynamics, the conserved variables are $(M,J,Q)$, with corresponding conjugate quantities temperature $T$, angular velocity $\Omega$, and electric potential $\Phi$ \cite{Davies1977}. Like pure fluid thermodynamics, black hole thermodynamics has well established notions of entropy, and zeroth, first, and second laws $(0,1,2)$ of thermodynamics. The third law $(3)$ of thermodynamics, however, is not obeyed in Kerr-Newman black hole thermodynamics since the entropy does not go to zero at zero temperature.

\begin{table}
\centering
\begin{tabular}{l|c|c}
\hline					& \,\,Pure fluid\,\,	& \,\,Kerr-Newman		\\
\hline
Conserved variables		& $(U, N, V)$		&$(M, J, Q)$			\\
Conjugate variables			& $(T,\mu,-p)$		& $(T,\Omega,\Phi)$		\\
Entropy?					& Yes			& Yes				\\
Thermodynamic laws (0,1,2)?	& Yes			& Yes				\\
Third law (3)?				& Yes			& No		                            \\
Extensive?				& Yes			& No					\\
Thermodynamically stable?	& Yes			& No		                            \\
Statistical mechanics?		& Yes			& Unclear				\\
\hline
\end{tabular}
\caption{Comparison between pure fluid thermodynamics and Kerr-Newman black hole thermodynamics.}
\label{tab:5}
\end{table}

\par
A clear difference between fluid and black hole thermodynamics is that black hole thermodynamics is not extensive \cite{Landsberg1980}. Namely, we cannot scale the mass of the black hole up in such a way as to leave all of the conjugate variables fixed. However, this point poses few difficulties for the black hole thermodynamic fluctuation theory employed here.

\par
Significant is the frequent absence of black hole thermodynamic stability. One manifestation of this are negative heat capacities, which are a fixture of gravitational thermodynamic problems. A black hole lacking thermodynamic stability cannot reach thermodynamic equilibrium with its environment, a significant deficit for the physical interpretation of any quantity, such as $R$, coming from thermodynamic fluctuation theory. The Kerr-Newman black hole thermodynamics is not stable for any set of values of $(M,J,Q)$ \cite{Tranah1980}. However, stable black hole cases do exist. These result on either restricting the number of fluctuating variables, adding an AdS background, or altering the assumptions about the black hole's topology. Stable thermodynamic cases get most of the attention in this talk.

\par
There is no consensus on the question of the correct microstructure supporting black hole thermodynamics. String theorists have attempted to calculate such microstructures, particularly for near extremal black holes, starting with Strominger and Vafa \cite{Strominger1996}; see Bellucci and Tiwari \cite{Bellucci2010,Bellucci2012} and Wei, et al., \cite{Wei2013} for recent references.\footnote{If a paper starts with a spacetime metric, and calculates the thermodynamic from the area of the event horizon, it is a general relativistic solution. If the paper starts with a Lagrangian, and a quantum action then it is beyond the scope of my talk.} In string theory calculations, the microscopic model is always explicit. By contrast, for general relativity solutions there is no evident microscopic foundation, and I direct my efforts to these in this talk.

\subsection{Laws of black hole thermodynamics}
\label{sec:43}

The Bekenstein-Hawking area law \cite{Bekenstein1974,Hawking1976} sets the black hole entropy $S_{BH}$ proportional to the area $A$ of the event horizon:

\begin{equation} \frac{S_{BH}}{k_B} = \frac{1}{4}\left(\frac{A}{L_p^2}\right),\label{110}\end{equation}

\noindent where

\begin{equation} L_p = \sqrt{\frac{\hbar G}{c^3}} \label{120}\end{equation}

\noindent is the Planck length. Here, $\hbar$ is Planck's constant divided by $2\pi$, $G$ is the universal gravitation constant, and $c$ is the speed of light. The area $A$ may be calculated in terms of the conserved variables, given a black hole solution from general relativity. Such a calculation yields the full black hole thermodynamics. For example, here is the formula for $A$ for the Kerr-Newman black hole \cite{Smarr1973}:

\begin{equation} A = 4\pi\left(2M^2-Q^2+2\sqrt{M^4-J^2-M^2 Q^2}\right).\label{130}\end{equation}

\par The black hole entropy may be added to the ordinary entropy $S_o$ to get the total entropy of the universe:

\begin{equation}S_{universe}=S_{BH} + S_o. \label{140}\end{equation}

\noindent We generalize the second law of thermodynamics in the obvious way, that in any process starting from some initial state and going to some final state:

\begin{equation}\Delta S_{universe}\ge 0. \label{150}\end{equation}

\par
Drop now the subscript ``BH'' ($S_{BH}\rightarrow S$), and turn to the first law of black hole thermodynamics. Writing $M=M(S, J, Q)$ leads to

\begin{equation} dM=T dS+\Omega dJ+\Phi dQ,\label{160}\end{equation}

\noindent where we define the temperature

\begin{equation}T=\left(\frac{\partial M}{\partial S}\right)_{J,Q},\label{170}\end{equation}

\noindent the angular velocity

\begin{equation}\Omega=\left(\frac{\partial M}{\partial J}\right)_{S,Q},\label{180}\end{equation}

\noindent and the electric potential

\begin{equation}\Phi=\left(\frac{\partial M}{\partial Q}\right)_{S,J}.\label{190}\end{equation}

\noindent The first law of black hole thermodynamics Eq. (\ref{160}) expresses the change in black hole energy $dM$ to mechanical work terms, $\Omega dJ$ and $\Phi dQ$, and a heat term $T dS$.

\par
Also essential is the $0$'th law of black hole thermodynamics, which equates $T$ to the effective surface tension of the event horizon. Calculations show this quantity to be constant over the event horizon, resulting in a unique value for the black hole temperature $T$. $\Omega$ and $\Phi$ are similarly constant over the event horizon \cite{Smarr1973}.

\par
Let me make one more observation about the correspondences in Table \ref{tab:5} before discussing black hole thermodynamic curvature. While there are natural correspondences between fluid and black hole energy, temperature, entropy, and (I argue) thermodynamic curvature $R$, there is always uncertainty in making correspondences among other thermodynamic variables. For example, if we have some fluid critical point property, say a divergence in the heat capacity at constant volume, one could not naturally say how this property translates to black hole thermodynamics. This point will be discussed further below in connection with black hole phase transitions.

\subsection{Black hole thermodynamic curvature $R$}
\label{sec:44}

\par
Black hole thermodynamics leads naturally to corresponding rules for black hole thermodynamic fluctuations, described by an information metric \cite{Ruppeiner2007,Aman2003,Pavon1983, Pavon1985} of the type in Eq. (\ref{20}). In conserved independent coordinates $(x^1,x^2,x^3,\cdots)=(X^1,X^2,X^3,\cdots)=(M,J,Q,\cdots)$, the thermodynamic metric for black hole fluctuations is (in appropriate units)

\begin{equation} g_{\alpha\beta} = -\frac{\partial^2 S}{\partial X^\alpha\partial X^\beta},\label{200}\end{equation}

\noindent where $S$ is the black hole entropy. The form of the thermodynamic metric in Eq. (\ref{200}) requires us to know $S=S(X^1,X^2,X^3,\cdots)$. Frequently, however, we know instead $M=M(Y^1,Y^2,Y^3,\cdots)$, where $(Y^1,Y^2,Y^3,\cdots)=(S,J,Q,\cdots)$. In this event, simplification results on writing the thermodynamic metric in the Weinhold energy form, with an additional prefactor $1/T$ \cite{Ruppeiner1995, Salamon1984}:

\begin{equation} g_{\alpha\beta} = \frac{1}{T}\frac{\partial^2 M}{\partial Y^\alpha\partial Y^\beta}.\label{204}\end{equation}

\noindent No matter how the thermodynamic metric is written, however, we will get the same value for $R$ for a given thermodynamic state, since $R$ is a thermodynamic invariant.\footnote{The line element Eq. (\ref{20}) transforms as a scalar, since probability is a scalar quantity. Hence, the metric elements $g_{\alpha\beta}$ transform as the elements of a second-rank tensor, which the relation between Eqs. (\ref{200}) and (\ref{204}) satisfies. The resulting thermodynamic curvature $R$ transforms as a scalar. These transformation properties hold under all coordinate transformations, including those resulting from Legendre transformations. This is the case in both ordinary and black hole thermodynamics, despite erroneous claims to the contrary \cite{Quevedo2008}.}

\par
Thermodynamic fluctuation metrics must be positive definite for thermodynamic stability. With two independent fluctuating variables, this requires {\it three} conditions:

\begin{equation} g_{11}>0,\label{206}\end{equation}

\begin{equation} g_{22}>0,\label{207}\end{equation}

\noindent and

\begin{equation} g_{11} g_{22} - g_{12}^2 > 0. \label{208}\end{equation}

\par
Pioneering papers introducing thermodynamic curvature $R$ into the black hole arena are \cite{Aman2003,Ferrara1997, Cai1999}. In particular, \r{A}man and Pidokrajt \cite{Aman2003} first evaluated $R$ for several solutions from general relativity. For nondiagonal thermodynamic metrics with $n=2$, such as those in Eqs. (\ref{200}) and (\ref{204}):

\begin{equation} \begin{array}{lr} {\displaystyle R= -\frac{1}{\sqrt{g}} \left[ \frac{\partial}{\partial x^1}\left(\frac{g_{12}}{g_{11}\sqrt{g}}\frac{\partial g_{11}}{\partial x^2}-\frac{1}{\sqrt{g}}\frac{\partial g_{22}}{\partial x^1}\right) \right. } \\  \hspace{3.6cm} + {\displaystyle \left. \frac{\partial}{\partial x^2}\left(\frac{2}{\sqrt{g}} \frac{\partial g_{12}}{\partial x^1} -\frac{1}{\sqrt{g}}\frac{\partial g_{11}}{\partial x^2}-\frac{g_{12}}{g_{11}\sqrt{g}}\frac{\partial g_{11}}{\partial x^1}\right)\right],}  \end{array} \label{210}\end{equation}

\noindent where 

\begin{equation} g=g_{11}g_{22}-g_{12}^2. \label{212}\end{equation}

\par
\begin{figure}
\centering
\includegraphics[height=4cm]{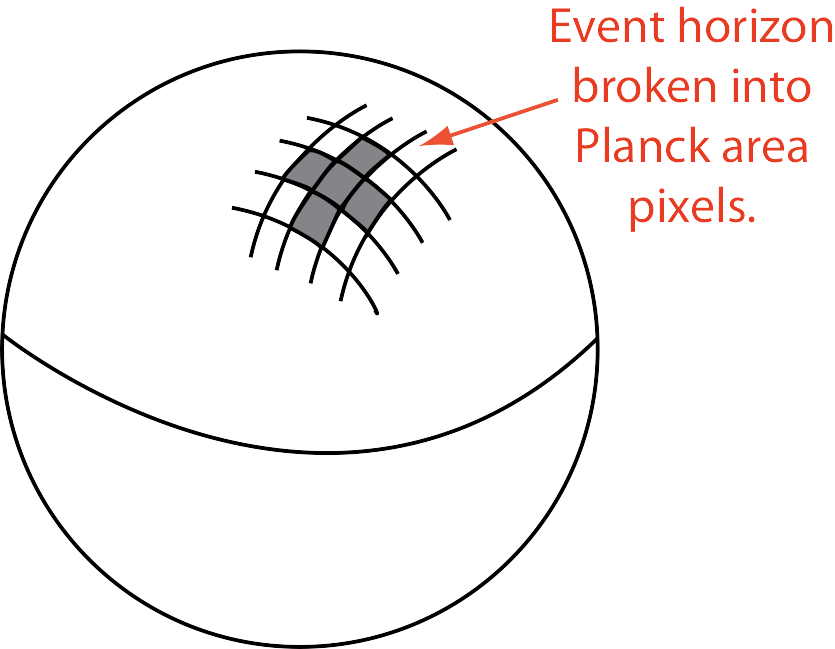}
\caption{The event horizon broken up into Planck area pixels. The dark pixels are portrayed as correlated. I propose that $|R|$ measures the average number of correlated pixels.}
\label{fig:8}
\end{figure}

But what is the physical interpretation of the black hole thermodynamic curvature $R$? In my view there is only one rational way to approach this question, and that is to follow the ideas developed in ordinary thermodynamics. It has been argued \cite{Ruppeiner2008} that the natural units of the thermodynamic curvature are the square of the Planck length $L_p^2$. Figure \ref{fig:8} shows the event horizon broken up into Planck area pixels. Perhaps $|R|$ measures the correlation between fluctuating Planck length pixels? Since I bring no microscopic theory of black holes into play in this talk, I have no direct evidence for such a conjecture. But, by analogy with the case in ordinary thermodynamics, how else could we interpret the black hole thermodynamic curvature?

\par
The picture in Fig. \ref{fig:8} assumes that all the black hole statistical activity takes place on the two-dimensional event horizon. This assumption is an element of the black hole membrane paradigm \cite{Thorne1986}. The motivation of the membrane paradigm is that if you cannot in principle know what is going on inside the black hole, then assume that all of the interesting stuff must be happening on the event horizon. One element of this idea is that if we are going to associate black hole statistics with some familiar model in statistical mechanics, then perhaps we should look most closely at two-dimensional models.

\subsection{Solutions from general relativity}
\label{sec:45}

\par
The thermodynamic curvature $R$ for black holes has been worked out for a number of systems, and I make no attempt to be complete in my reporting below. Rather, I present some thoughts about how results from various general relativity solutions might be compared with one another, and to solutions from ordinary thermodynamics. In Tables \ref{tab:6} and \ref{tab:7}, I consider only thermodynamic states with $S>0$, $M>0$, and $T>0$. Within this physical range of variables, the solutions divide into two categories, those for which there are no regimes satisfying thermodynamic stability (Eqs. (\ref{206}), (\ref{207}), and (\ref{208})), and those for which there are such regimes. In either category, $R$ can be readily worked out from Eq. (\ref{210}); it is real in all the cases I calculated. However, the physical interpretation I have presented for $R$ for ordinary thermodynamics is based on fluctuation theory, and this assumes thermodynamic stability. I key on the stable cases below.

\par
Table \ref{tab:6} shows results for $R$ for several general relativity solutions having no stable thermodynamic states. Tabulated are the dimension (spatial $+$ time), the fluctuating conserved variables, whether or not there are regimes of thermodynamic stability (no cases in Table \ref{tab:6}), the sign of $R$ (or an indication ``0'' if $R$ is identically zero), whether or not there are places where the sign of $R$ changes through zero, and whether or not there are divergences $|R|\to\infty$. Of the older solutions: Reissner-Nordstr$\ddot{\mbox{o}}$m, Kerr, and Kerr-Newman, none are thermodynamically stable for any thermodynamic state. Also, not thermodynamically stable are the two solutions listed with a higher dimension $=4+1$. Some of the older solutions have been worked out in higher dimensions, but with no reports of thermodynamically stable cases \cite{Aman2006}.

\begin{table}
\begin{center}
\begin{tabular}{l|c|c|c|c|c|c}
\hline
Name of solution										& Dimension		& Variables		& Stable	& $R$ sign	& $R=0$	& $|R|$ divergence	\\
\hline
Reissner-Nordstr$\ddot{\mbox{o}}$m\cite{Aman2003}			& $3 + 1$			& $(M, Q)$		& none	& $0$		& -		& none			\\
Kerr \cite{Aman2003}									& $3 + 1$			& $(M, J)$			& none	& $+$		& no		& extremal		\\
Kerr-Newman \cite{Ruppeiner2008,Aman2003,Mirza2007}		& $3 + 1$			& $(M, J, Q)$		& none	& $+$		& no		& extremal		\\
Black hole \cite{Arcioni2005}								& $4 + 1$			& $(M, J)$			& none	& $+$		& no		& extremal		\\
Small black ring \cite{Arcioni2005}							& $4 + 1$			& $(M, J)$			& none	& $\pm$		& yes	& ext+crit line		\\
\hline
\end{tabular}
\end{center}
\caption{The thermodynamic curvature $R$ for black hole solutions from general relativity. The solutions shown here have no regimes of thermodynamic stability. ``Extremal'' denotes a curve in the space of variables with $T=0$, and ``crit'' denotes a critical line with $T\ne 0$, along which $|R|$ diverges.}
\label{tab:6}
\end{table}

\par
Table \ref{tab:7} shows results for $R$ for several general relativity solutions with ``some'' or ``all'' states thermodynamically stable. Thermodynamic stability results on either adding an AdS background, restricting the number of fluctuating variables, or altering the assumptions about the black hole's topology.

\begin{table}
\begin{center}
\begin{tabular}{l|c|c|c|c|c|c}
\hline
Name of solution										& Dimension		& Variables		& Stable	& $R$ sign	& $R=0$	& $|R|$ divergence	\\
\hline
BTZ \cite{Aman2003,Cai1999}								& $2 + 1$			& $(M, J)$			& all		& $0$ 		& 0		& none			\\
RN-AdS \cite{Aman2003,Shen2007,Sahay2010b,Niu2012}		& $3 + 1$			& $(M, Q)$		& some	& $\pm$		& yes	& ext+crit line		\\
K-AdS \cite{Sahay2010b,Banerjee2011,Tsai2012}				& $3 + 1$			& $(M, J)$			& some	& $-$		& no		& critical line		\\
Restricted KN \cite{Ruppeiner2008,Ruppeiner2007}			& $3 + 1$			& $(J, Q)$			& all		& $+$		& no		& extremal		\\
Large black ring \cite{Arcioni2005}							& $4 + 1$			& $(M, J)$			& all		& $-$		& no		& ext+crit line		\\
\hline
\end{tabular}
\end{center}
\caption{The thermodynamic curvature $R$ for black hole solutions from general relativity. These solutions all have at least some thermodynamically stable regimes. The characterization of $R$ is based only on states in the stable regime. ``Extremal'' denotes a curve in the space of variables with $T=0$, and ``crit'' denote a critical line with $T\ne 0$ along which $|R|$ diverges.}
\label{tab:7}
\end{table}

\par
\begin{figure}
\begin{minipage}[b]{0.5\linewidth}
\includegraphics[width=2.0in]{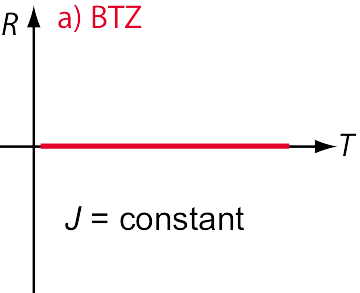}
\end{minipage}
\hspace{0.2cm}
\begin{minipage}[b]{0.5\linewidth}
\includegraphics[width=2.0in]{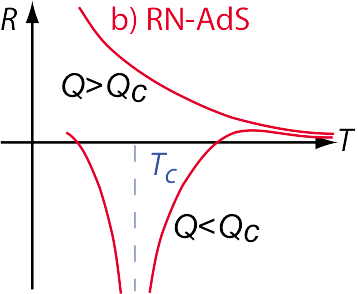}
\end{minipage}
\vspace{0.4cm}
\begin{minipage}[b]{0.5\linewidth}
\includegraphics[width=2.0in]{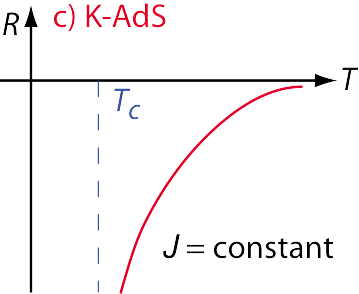}
\end{minipage}
\hspace{0.2cm}
\begin{minipage}[b]{0.5\linewidth}
\includegraphics[width=2.0in]{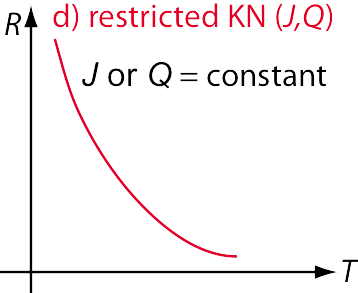}
\end{minipage}
\vspace{0.4cm}
\begin{minipage}[b]{0.5\linewidth}
\includegraphics[width=2.0in]{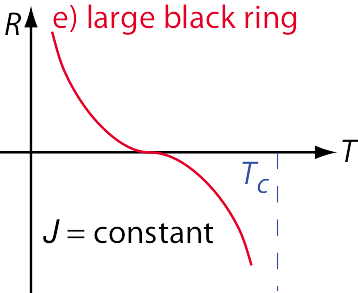}
\end{minipage}
\caption{Schematic graphs for $R$ for thermodynamically stable general relativistic black hole solutions: (a) the BTZ solution, with $R=0$; (b) the RN-AdS solution, with $R$ diverging to positive infinity at the extremal curve for $Q>Q_c$, and with $R$ diverging to negative infinity, at temperature $T_c>0$, along the critical line for $Q<Q_c$; (c) the K-AdS solution, with $R$ diverging to negative infinity at the critical line, (d) the restricted KN $(J,Q)$ solution, with $R$ diverging to positive infinity at the extremal limit; (e) the large black ring solution, with $R$ diverging to positive infinity at the extremal curve, and with $R$ diverging to negative infinity along the critical line.}
\label{fig:9}
\end{figure}

Black holes in an AdS background have significant regimes of thermodynamic stability. The simplest member of this category is the BTZ black hole, which is thermodynamically stable for all of its states, and has identically zero $R$. This behavior is shown schematically in Figure \ref{fig:9}a, and it resembles the behavior for the ideal gas in Fig. \ref{fig:4}a.

\par
In the thermodynamically stable regime, Reissner-Nordstr$\ddot{\mbox{o}}$m-AdS black holes have an extremal curve $T=0$, as well as a line of critical points where $|R|$ diverges to infinity. This critical line obtains for $Q<Q_c$, where the critical value $Q_c$ depends on the cosmological constant. For a fixed $Q>Q_c$, as we reduce $T$ from a large value, $R$ diverges to positive infinity at the extremal curve. However, for fixed $Q<Q_c$, as we reduce $T$ from a large value, $R$ diverges to {\it negative} infinity along the critical line. As $T$ is decreased further, we enter a thermodynamically unstable regime followed by a stable regime where $R$ increases.

\par
The general black hole thermodynamic behavior for RN-AdS has been associated with a phase transition analogous to a van der Waals model by Chamblin, et al., \cite{Chamblin1999a,Chamblin1999b}. A number of researchers have calculated $R$ for this case \cite{Aman2003,Shen2007,Sahay2010b,Niu2012}. The behavior of $R$ for RN-AdS is shown schematically in Fig. \ref{fig:9}b. For $Q>Q_c$ the behavior of $R$ resembles that in the Fermi gas, shown in Fig. \ref{fig:4}i. For $Q<Q_c$, $R$ resembles the critical point behavior in Fig. \ref{fig:4}e. This correspondence is certainly consistent with the association with the van der Waals model. I add that the $Q<Q_c$ curve in Fig. \ref{fig:9}b has a bump resembling the one in Fig. \ref{fig:4}g.

\par
Kerr-AdS black holes have no extremal curve in the thermodynamically stable regime. However, a critical line depending on the cosmological constant bounds the thermodynamically stable regime at low $T$. Along this critical line, $R$ diverges to negative infinity. For the Kerr-AdS black hole thermodynamics, $R$ is always negative. Fig. \ref{fig:9}c sketches $R$ as $T$ is decreased from a large value at constant $J$. The sketch resembles the critical point behavior in Fig. \ref{fig:4}e. However, once we cross the critical line, there are no more thermodynamically stable regimes. Banerjee, et al., \cite{Banerjee2011, Banerjee2012} have discussed phase transitions in AdS black holes using the Ehrenfest relations, with special attention to the orders of the phase transitions.

\par
Special thermodynamically stable cases result from the Kerr-Newman solution when we fix one of the three parameters $(M,J,Q)$, and allow the other two to fluctuate. This restriction is not just a mathematical convenience; it has a physical basis. For example, consider adding an electron to the black hole, and calculate the contribution of each of changing $(M,J,Q)$ to the change in the total entropy. We expect one of $(M,J,Q)$ to contribute least to the changing entropy, and if it contributes much less, we could just ignore the change in that parameter, and let the other two parameters fluctuate. For a black hole with mass on the order of the Planck mass (a quantum black hole), contributions to the changing entropy from the electron mass are hugely less important to the change in total entropy than the changes resulting from its $(J,Q)$. This restricted KN $(J,Q)$ solution has some highly desirable properties, as Table \ref{tab:7} shows. This solution is sketched in Fig. \ref{fig:9}d. In addition, there are some detailed analogies to the 2D ideal Fermi gas in the extremal limit, which may be interesting.

\par
The large black ring solution also has significant regimes of stability, bounded by an extremal curve and a critical line. $R$ diverges to positive infinity at the extremal curve, and to negative infinity along the critical line.

\par
A few patterns present themselves for the thermodynamically stable general relativity solutions considered in this section. In all cases, the divergence of $R$ at the extremal curve is to positive infinity, resembling in this sense the divergence for the ideal Fermi gasses from ordinary thermodynamics. Where there are critical lines (with $|R|$ diverging with $T\ne 0$), the divergence of $R$ is to negative infinity, resembling the critical point divergences in ordinary thermodynamics. But I have considered too few cases here to assert with any confidence that these patterns are general. Further study is obviously necessary.

\subsection{Discussion of ``inconsistencies''}
\label{sec:46}

\par
Much debated in black hole thermodynamics has been the possibility raised by Davies \cite{Davies1977} that the curve of diverging heat capacity $C_{J,\,Q} = T(\partial S/\partial T)_{J,\,Q}$ in the Kerr-Newman black hole solution corresponds to a phase transition. Diverging heat capacities are a feature of second-order phase transitions in ordinary fluid and spin systems, so Davies' association would appear logical.

\par
\begin{figure}
\centering
\includegraphics[height=5cm]{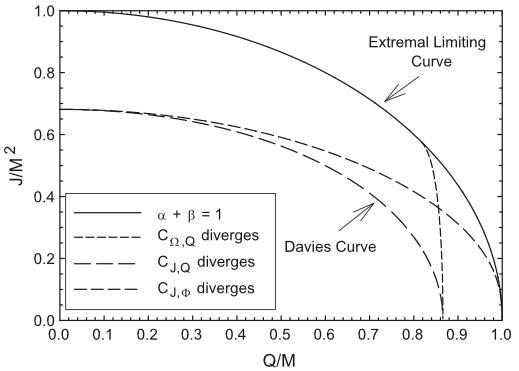}
\caption{Characteristic curves for the Kerr-Newman black hole. The curve along which $C_{J,Q}$ diverges is the Davies curve. $R$ diverges at the extremal limit and along curves corresponding to a change of thermodynamic stability, which have diverging $C_{J,\Phi}$ and $C_{\Omega,Q}$. $R$ does not diverge along the Davies curve. Here $\alpha=J^2/M^4$ and $\beta=Q^2/M^2$.}
\label{fig:10}
\end{figure}

Closer examination, however, raises some questions about Davies' correspondence. First, an ordinary thermodynamic system generally has at its foundation some {\it known} microscopic model. Such a model offers direct insight not only into the character of the thermodynamic variables, but into the microscopic signatures of any thermodynamic anomaly. In the absence of a known microscopic model we have difficulty answering basic questions. If some heat capacity diverges, how could we be sure that we have not just made an inappropriate choice of thermodynamic variables, which reveals infinities with no really fundamental significance? What do we make of curves in thermodynamic state space where one heat capacity diverges, but the other heat capacities stay regular? What if various heat capacities diverge along different curves, as happens in the Kerr-Newman black hole \cite{Ruppeiner2008,Tranah1980}, as in Figure \ref{fig:10}. Which curve corresponds to a true phase transition? One of them? All of them? Perhaps it is safer to associate curves of diverging $R$ with black hole phase transitions. $R$ has a unique status in identifying microscopic order from thermodynamics. and ordering at the microscopic level is at the foundation of phase transitions.

\par
Black hole solutions with $R$ identically zero, of which Tables \ref{tab:6} and \ref{tab:7} each have one, have also given rise to debate; see \cite{Medved2008} for a review. If $R$ measures in some sense the range of interactions, then one might expect $|R|$ to always be large for black hole thermodynamics, reflecting the concentrated gravitational forces present in these objects. But such reasoning need not obtain. In a classical black hole, the gravitating particles have collapsed to a central singularity, shrinking the interactions between them to zero volume. The statistics underlying the thermodynamics might reside on the event horizon, where unknown constituents might interact with each other by forces perhaps not gravitational. In this scenario, gravity might merely be a nonstatistical force holding the assembly together, and a result $R = 0$, where the unknown constituents move independently of each other, would make perfect sense.

\section{Conclusions}
\label{sec:6}

What are black holes made out of? This question has not been answered here. However, one way to address this question is by following an agenda of matching the statistical mechanics of known microscopic models to black hole thermodynamic solutions from general relativity, or other theories of gravity. I hope that I have convinced the audience that the thermodynamic curvature $R$ has a contribution to make to this game.

\par
I have given a broad survey of thermodynamic curvature $R$, one spanning results in fluids and solids, spin systems, and black hole thermodynamics. $R$ results from the unique thermodynamic information metric giving thermodynamic fluctuations. $R$ has a unique status in thermodynamics as being a geometric invariant, the same for any given thermodynamic state no matter what coordinates we calculate in. In ordinary thermodynamics, the sign of $R$ indicates the character of microscopic interactions, and $|R|$ indicates the average size of organized fluctuations. Although I have given no direct evidence that this interpretation holds for black hole thermodynamics, if we believe in the broad generality of thermodynamic principles, this interpretation of $R$ should transcend specific scenarios.

\par
Most incomplete in this talk has been the presentation of spin systems. Frustration in spin systems may be necessary as a way to deal with the frequent failure of the third law of thermodynamics for black holes. Missing entirely from this talk have been results on string theory models, which were simply beyond reach of the speaker. These may ultimately yield the best picture of what is going on in the black hole.

\section{Acknowledgments}

I thank George Skestos for travel support, and I thank the conference organizer Stefano Bellucci of INFN for giving me the opportunity to present my work.

\end{document}